\documentclass[final,3p,times]{elsarticle}

\usepackage{amssymb}
\usepackage{comment}
\usepackage{braket}

\def\im{\mathrm i}

\usepackage{color,ulem}

\journal{Physica E}

\begin{document}

\begin{frontmatter}



\title{Finite Size Effects of the Surface States in a Lattice Model of Topological Insulator}


\author{Kazuto Ebihara, Keiji Yada, Ai Yamakage, and Yukio Tanaka}

\address{Department of Applied Physics, Nagoya University, 464-8603, Japan}

\begin{abstract}
Energy gap and wave function in thin films of topological insulator is studied, based on tight--binding model.
It is revealed that 
thickness dependence of the magnitude of energy gap
is
composed of damping and oscillation.
The damped behavior originates from the presence of gapless surface Dirac cone in the infinite thickness limit.
On the other hand,
the oscillatory behavior stems from 
electronic properties 
in the thin thickness limit.
\end{abstract}


\end{frontmatter}


\section{Introduction}
\label{introduction}

Recently topological insulator has attracted much attention \cite{hasan10,qi10_QSH,qi10_topo,moore10,hasan11}.
Topological insulator has been firstly predicted 
in a graphene with spin-orbit interaction 
\cite{kane05QSH}, which consists of two copies of quantum Hall system, and shows a quantized spin Hall effect if $z$--component of spin of electrons is conserved.
Generally speaking, the present non--trivial system is characterized by $\mathbb Z_2$--index introduced by Fu, Kane, and Mele \cite{kane05Z2,fu06}, and has a gapless helical edge mode where spin current protected by time--reversal symmetry  flows spontaneously.
However, quantum spin Hall (QSH) phase of graphene has not been observed experimentally since the spin--orbit interaction that is the driving force of topological insulator is much small.
After that,
HgTe/HgCdTe quantum well has been theoretically proposed as a candidate of  two--dimensional topological insulator  \cite{bernevig06}, and confirmed experimentally \cite{koenig07,roth09,brune11}.
Topological insulators have been realized in three dimensional systems \cite{fu07_3D,fu07_inversion,moore07}, {\it e.g.}, Bi$_{1-x}$Sb$_{x}$ alloy \cite{hsieh08}, the binary compounds Bi$_2$Se$_3$, Bi$_2$Te$_3$ \cite{zhang09,liu10_model,xia09,chen09}, and Tl--based ternary compound TlBiSe$_3$ \cite{yan10,lin10,sato10}.
Furthermore,
the quaternary compounds have also been theoretically predicted \cite{chen11,wang11arXiv}.
All of these systems have a single helical Dirac cone on the surface.

Nowadays, many exotic quantum phenomena are expected originating from surface states of three--dimensional topological insulators \cite{hasan10,qi10_QSH,TYN09,Linder10a,Linder10b2,Yokoyama}.
However, in the actual systems,
 sufficient amount of carriers remain in the bulk due to 
the difficulty of fabrication of samples \cite{hyde74}.
Then
 the system becomes metallic and it is difficult to classify physical properties specific to surface Dirac cone \cite{chen09,taskin09,checkelsky09}.
To resolve this problem, 
several approaches, \textit{e.g.,} chemical doping and surface adsorption \cite{hsieh09} have been performed.
The another new approach to control the carrier is 
to fabricate high quality thin films, where carrier control by gating is possible 
\cite{chen10PRL,chen10PRB}.
But 
thin films may have the different electronic states from that of the bulk.
Especially, the surface states have an energy gap due to the hybridization between the Dirac cones on top and bottom on the film induced by finite--size effect.

Based on above backgrounds, 
experimental studies of thin films of topological insulators have started \cite{zhang10, sakamoto10, hirahara10}.
Besides this,
there have been many theoretical studies based on
continuous models \cite{linder09, lu10, shan10}, first principle calculations \cite{liu10_oscillatory, park10, yazyev10, jin11, chang11}, and tight--binding model calculation \cite{liu10_oscillatory}.
%
%
%
Although
continuous model is simple,
it is
valid only for the long wavelength and low--energy limits.
First principle calculation gives detailed electronic states of thin film.
But
it is difficult to analyze
complicated phenomena, \textit{e.g.}, 
transport properties, disorder effects, and quantum many--body problems.
%
%
%
On the other hand,
tight--binding approach
is useful to  calculate these interesting phenomena numerically, because many--body interaction and impurity effects are easily taken into account.
However, electronic properties of thin film of topological insulator have not been fully studied based on tight--binding model.

In the present paper, we study 
electronic properties of thin film of
Bi$_2$Se$_3$ based on 
a tight--binding model focusing on the
film--thickness dependencies of energy gap and surface states.
It is revealed that
the magnitude of energy gap is seriously influenced by material parameters.
The paper is organized as follows. 
In section 2, we introduce a tight--binding model based on the Hamiltonian proposed by Refs. \cite{zhang09,liu10_model}.
In section 3, we calculate 
energy spectrum of thin film for various number of quintuple layers
by changing material parameters.
In section 4, we conclude our results.

\section{Model}
\label{model}

We use the effective model derived in Refs. \cite{zhang09,liu10_model}
 as
\begin{eqnarray}
	H(\mathbf k) =
	\mathcal{E}({\bf k}) + \left(
	\begin{array}{cccc}
	{\cal M}({\bf k}) & 0 & B_0 k_z & A_0 k_- \\
	0 & {\cal M}({\bf k}) & A_0 k_+ & -B_0 k_z \\
	B_0 k_z & A_0 k_-&-{\cal M}({\bf k})&0 \\
	A_0 k_+ & -B_0 k_z & 0 & -{\cal M}({\bf k}) \\
	\end{array}
	\right),
\end{eqnarray}
with
\begin{eqnarray}
	\mathcal{E}({\bf k}) &=& C_0+C_1k_z^2+C_2(k_x^2+k_y^2), \\
	{\cal M}({\bf k}) &=& M_0+M_1k_z^2+M_2(k_x^2+k_y^2),
\end{eqnarray}
where $k_\pm=k_x\pm\im k_y$, and the base is taken as
$(|+,\uparrow\rangle, | +, \downarrow \rangle, |- , \uparrow \rangle,
| -, \downarrow \rangle)$, in which $\pm$ and $\uparrow(\downarrow)$
denote the parity eigenvalue and spin respectively.
Let us introduce the lattice model only with nearest neighbor hoppings in a tetragonal lattice with substitution as
\begin{eqnarray}
 k_i a_i \to \sin k_i a_i, \quad (k_ia_i)^2 \to 2 (1-\cos k_i a_i).
\end{eqnarray}
As a result,
the bulk Hamiltonian is derived as
\begin{eqnarray}
H(\mathbf k) =
	\tilde\mathcal{E}({\bf k}) + \left(
	\begin{array}{cccc}
	\tilde {\cal M}({\bf k}) & 0 & B_0 \sin k_zc & \bar A_- \\
	0 & \tilde{\cal M}({\bf k}) & \bar A_+ & -B_0 \sin k_zc \\
	B_0 \sin k_zc & \bar A_- &- \tilde{\cal M}({\bf k})&0 \\
	\bar A_+ & -B_0 \sin k_zc & 0 & -\tilde{\cal M}({\bf k}) \\
	\end{array}
	\right),
	\label{Hbulk}
\end{eqnarray}
where
\begin{eqnarray}
\tilde\mathcal E(k_x,k_y) &=& \bar C_0 + 2 \bar C_1(1- \cos k_z c ) + 2\bar C_2(2-\cos k_xa-\cos k_ya) ,\\
\tilde {\cal M}({\bf k}) &=& \bar M_0 + 2 \bar M_1 (1- \cos k_z c ) + 2 \bar M_2 (2 - \cos k_x a - \cos k_y a),
\\
\bar A_{\pm}(k_x,k_y) &=& 
\bar A_0 (\sin k_xa \pm \mathrm i \sin k_ya),
\end{eqnarray}
with $a_i$ ($ a \equiv a_x=a_y, \, c \equiv a_z$) being the lattice constant along $i(=x,y,z)$--direction.
The relation between the original parameters
and
those in the present model
is
\begin{eqnarray} 
\bar M_0 = M_0,  \bar C_0 = C_0,
\bar M_1 = M_1/c^2,  \bar C_1 = C_1/c^2,
\nonumber\\
\bar M_2 = M_2/a^2,  \bar C_2 = C_2/a^2,
\bar A_0 = A_0/a, \bar B_0 = B_0/c.
\end{eqnarray}
In the following,
we express
the present Hamiltonian in real space along $z$--direction 
perpendicular 
to the quintuple layers
to focus on the surface states.
Here, 
translational invariance is 
satisfied for the direction parallel to the quintuple layers
i.e., $x$-- and $y$--directions.
Then $k_x$ and $k_y$ are good quantum numbers.
We apply open boundary condition only along $z$--direction, {\it i.e.}, the system is regarded as a one--dimensional chain for fixed $(k_x,k_y)$.
This condition corresponds to (111) cleavage surface of actual ${\rm Bi_2Se_3}$ which is easily cleaved.
The corresponding Hamiltonian is given as follows,
\begin{eqnarray}
 H(k_x,k_y) &=& \sum_{n=1}^{N_z} 
	c^\dag_n(k_x,k_y) H_0(k_x,k_y) c_{n}(k_x,k_y)
  \nonumber\\ && \quad +
  \sum_{n=1}^{N_z-1}
	\left[
		c^\dag_n(k_x,k_y) H_1 c_{n+1}(k_x,k_y)
		+
		\mathrm{h.c.}
	\right],
 \label{tb}
\end{eqnarray}
 where $N_z$ denotes number of quintuple layers.
It is noted that a lattice point $n$ in the above Hamiltonian
corresponds to position of a 
quintuple layer in the actual crystal structure.
%
The on--site energy is given by
\begin{eqnarray}
	H_0(k_x,k_y) =  \left(
	\begin{array}{cccc}
	\bar{\cal M} + \bar{\cal E} & 0&0& \bar A_{-} \\
	0 & \bar{\cal M} + \bar{\cal E} & \bar A_+ &0\\
	0 & \bar A_- & -\bar{\cal M} + \bar{\cal E}  &0 \\
	\bar A_+ &0&0& -\bar{\cal M} + \bar{\cal E} 
	\end{array}
	\right),
 \label{H0}
\end{eqnarray}
with
\begin{eqnarray}
\bar\mathcal E(k_x,k_y) = \bar C_0 + 2 \bar C_1 + \bar C_2(2-\cos k_xa-\cos k_ya),\\
\bar\mathcal M(k_x,k_y) = \bar M_0 + 2 \bar M_1 + \bar M_2(2-\cos k_xa-\cos k_ya), 
\end{eqnarray}
and  the hopping between the nearest layers is as follows
\begin{eqnarray}
	H_{1} = \left(
	\begin{array}{cccc}
	-\bar M_1 - \bar C_1 & 0 &  \im \bar B_0/2 & 0 \\
	0 & - \bar M_1 - \bar C_1 & 0 & - \im \bar B_0/2 \\
	\im \bar B_0/2 & 0 & \bar M_1 - \bar C_1 & 0 \\
	0 & - \im \bar B_0/2 & 0 & \bar M_1 - \bar C_1 \\
	\end{array}
	\right).
\end{eqnarray}

Since the Hamiltonian $H(k_x,k_y)$ has an inversion symmetry, it follows that $[H,P]=0$, or equivalently $PH(k_x,k_y)P^{-1}=H(-k_x,-k_y)$, where the parity operator $P$ is defined by
\begin{eqnarray}
 P&=& \sum_{n=1}^{N_z} c_n^\dagger (k_x,k_y) {\rm diag}[1,1,-1,-1] c_{N_z+1-n}(k_x,k_y).
\end{eqnarray}
By using the parity operator $P$, we can derive the topological invariants $\nu$, which can be deduced from the parity of each pair of Kramers degenerate occupied energy band at the four time-reversal points at $\Gamma_\alpha$ ($\Gamma_1=(0,0)$, $\Gamma_2=(\pi,0)$, $\Gamma_3=(0,\pi)$, $\Gamma_4=(\pi,\pi)$,) in the Brillouin zone,
\begin{eqnarray}
(-1)^{\nu}&=&\prod_{\alpha=1}^{4} \prod_{m=1}^{N_z}\Braket{\phi_{2m}(\Gamma_\alpha)|P|\phi_{2m}(\Gamma_\alpha)},
\end{eqnarray}
where $\phi_m(\Gamma_\alpha)$ is the eigenvector of the Hamiltonian $H(\Gamma_\alpha)$, and $\Braket{\phi_{m}(\Gamma_\alpha)|P|\phi_{m}(\Gamma_\alpha)} (=\pm1)$ is the eigenvalue of parity operator $P$.

\section{Results and discussions}

We numerically obtain the eigenvalues and eigenvectors of bulk and surface states, diagonalizing the Hamiltonian given by eq. (\ref{tb}).
The value of parameters $\bar M_0, \bar M_2, \bar A_0, \bar C_0, \bar C_2$ are 
the same as in Ref. \cite{liu10_model} with using $a=4.14 \AA$.
The values of $\bar M_1$ and $\bar C_1$ are determined 
so that the eigen--energy at $Z$--point in Brillouin zone
coincides with that of first principle calculation in Ref. \cite{liu10_model}.
The value of $\bar{B_0}$ is chosen 
in order to fit 
the dispersion
along $\Gamma-Z$ line 
as well as possible. 
\begin{figure*}
\centering
\includegraphics{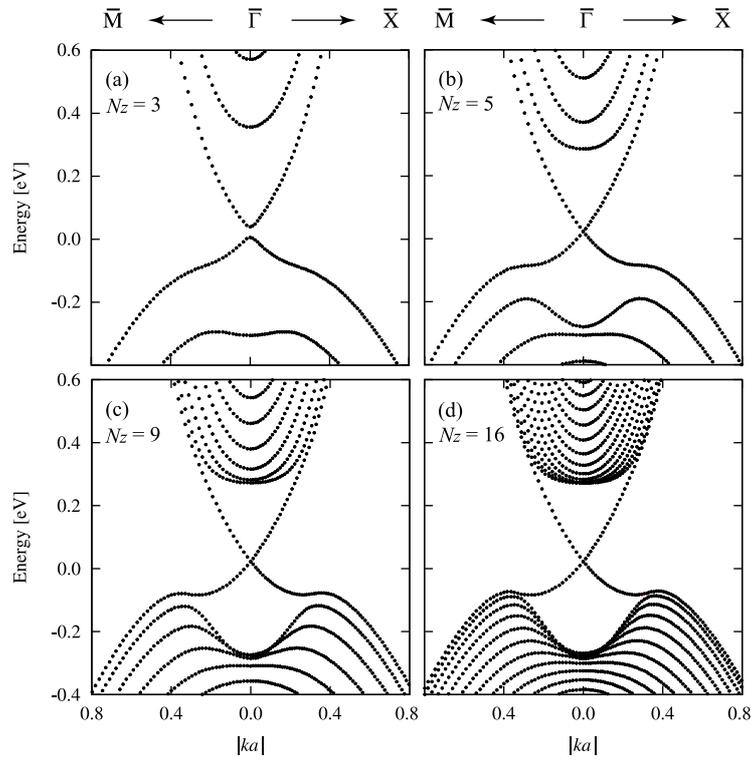}
\caption{Energy spectra of the bulk and surface states near $\bar{\Gamma}$-point in the slab geometry for $N_z= 3, 5,9$, and 16 quintuple layers.
The material parameters are set as $\bar M_0 = -0.28 \, \mathrm{eV}, \, \bar M_1=0.216 \, \mathrm{eV}, \, \bar M_2 = 2.60 \, \mathrm{eV}, \, \bar A_0 = 0.80 \, \mathrm{eV}, \, \bar B_0 = 0.32 \, \mathrm{eV}, \, \bar C_0 = -0.0083 \, \mathrm{eV}, \, \bar C_1 = 0.024 \, \mathrm{eV}, \, \bar C_2 = 1.77 \, \mathrm{eV}, \, a = 4.14 \, \AA, \,$ and $c= 9.55 \, \AA$.
}
\label{EV}
\end{figure*}
The indirect energy gap in the bulk Hamiltonian is located between $-0.071 \, {\rm eV}$  and $|\bar M_0 + \bar C_0| = 0.29 \, \rm eV$,
as derived from  eq. (\ref{Hbulk}).

Figure \ref{EV} shows the energy spectrum for a slab geometry in the cases of $N_z = 3,5,9,$ and 16.
We can clearly see that  eigenstates exist within the bulk energy gap.
These states can be regarded as surface states, which we can directly confirm from its density distribution localized in the vicinity of surface, as shown in Figure \ref{rho}.
The surface states have
a large magnitude of energy gap $E_{\rm g} \sim 0.033 \, \rm eV$
for $N_z=3$ 
since the two wave functions localized at the top and bottom surfaces overlap significantly. 
(see (a) in Figure \ref{EV}).
The magnitude of the present energy gap 
becomes small with the increase of $N_z$.
((b) and (c) in Fig. \ref{EV}.)
For $N_z=16$,  the resulting $E_{\rm g}$ is significantly reduced
to be $0.15 \times 10^{-6} \, \rm eV$
((d) in Figure \ref{EV}).
Moreover, 
it is noted
that the shape of valence subband depends on $N_z$.
For $N_z=3$ ((a) in Fig. \ref{EV}), there are two valence subbands in $-0.4 \, {\rm eV}< E < 0 {\rm eV}$. 
The upper subband consists mainly of surface states 
since it is located in the bulk energy gap.
The lower one, that is bulk energy band, is located at $\sim -0.3 \rm eV$.
The new subband appears between these two subbands at $\sim -0.2 \rm eV$ for $N_z=5$ ((b) in Fig. \ref{EV}). 
Simultaneously, 
valence subbands have 
a local minimum at $k=0$, 
and an indirect energy gap is generated.
For $N_z=9$ ((c) in Fig. \ref{EV})
there are much more subbands.
The energy bands
for $N_z=16$  as shown in (d) in Fig. \ref{EV} is almost similar to that of bulk three-dimensional topological insulator with $N_z \to \infty$.

\begin{figure}
\centering
\includegraphics[scale=1.]{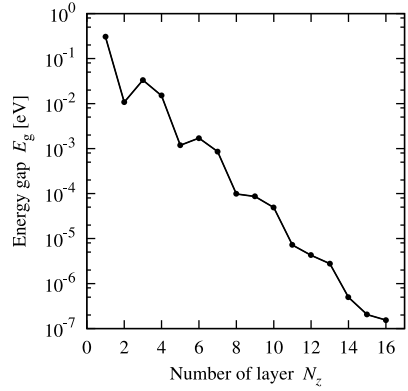}
\caption{The magnitude of energy gap $E_{\rm g}$ of the surface state as a function of $N_z$. The material parameters are the same as in Figure \ref{EV}.}
\label{gap}
\end{figure}

\begin{figure}
\centering
\includegraphics[scale=1.]{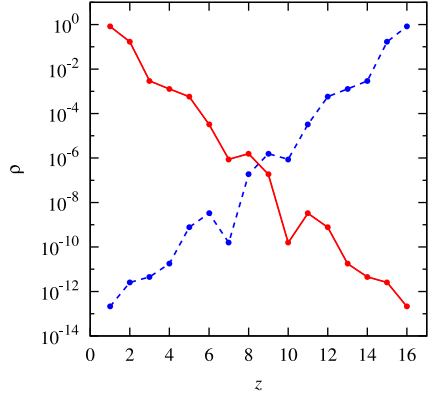}
\caption{Density distribution $\rho(z)$  of surface state at $(k_xa, k_ya) = (\pi/32,\pi/32)$ for $N_z=16$. There are two surface states located at the top and bottom in the system. $z$ denotes the position of layer. The guided lines are drawn for view-ability. The material parameters are the same as in Figure \ref{EV}.}
\label{rho}
\end{figure}

In Figure \ref{gap},
$N_z$ dependence of $E_{\rm g}$ is plotted.
Hamiltonian of the monolayer system with $N_z=1$ is given by $H_0(k_x, k_y)$ (see eq. (\ref{H0})), which is equivalent to that of HgTe/HgCdTe quantum well \cite{bernevig06},
and $E_{\rm g}$ is given by
$2 |\bar M_0 + 2 \bar M_1| (= 0.30 \mathrm{eV})$.
$E_{\rm g}$ for $N_z=2$ is also derived analytically as $|2 \bar M_1 - [{\bar B_0^2 + 4(\bar M_0 + 2 \bar M_1)^2}]^{1/2} | (=0.0094 \mathrm{eV})$.
For $N_z = 2,3$, and $4$,
$E_{\rm g}$ decreases roughly exponentially as a function of $N_z$, and becomes  $E_{\rm g} \sim 10^{-1} \, \rm eV$.
$E_{\rm g}$ becomes much smaller than room temperature for $N_z \geq 5$.
It is also noted that
$E_{\rm g}$ has an oscillatory behavior 
as a function $N_z$
whose period is almost $3$.
The similar behavior has been obtained
based on continuous models \cite{linder09, lu10, shan10} and first principle calculations \cite{liu10_oscillatory, park10, yazyev10, jin11, chang11}.

Next, we investigate the relation between the magnitude of energy gap and wave functions.
Figure \ref{rho} shows
the density distribution of surface states for $N_z=16$, which is defined by
\begin{eqnarray}
 \rho(z) = \left\langle  c^\dag_z(k_x,k_y) c_z(k_x,k_y) \right\rangle,
 \quad
 z = 1, \cdots, N_z,
\end{eqnarray}
where the expectation value is evaluated for the surface state with momentum $(k_xa,k_ya) = (\pi/32, \pi/32)$.
The solid (dashed) line denotes the density distribution of surface state located on the top $z=1$ (bottom $z=16$).
The density distribution decays exponentially with oscillation whose period is nearly 3 quintuple layers.
This period is almost the same as that of $N_z$ dependence of $E_{\rm g}$.
Since
the two wave functions located at the top and bottom surfaces oscillate spatially,
the resulting $E_{\rm g}$ due to overlap between them also oscillates as a function of $N_z$.

\begin{figure}
\centering
\includegraphics[scale=0.5]{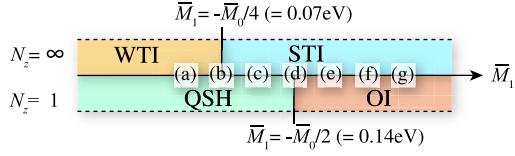}
\caption{Phase diagrams of topological insulator for bulk limit $N_z = \infty$ and for monolayer $N_z=1$. 
STI (WTI) denotes strong (weak) topological insulator 
for $N_z=\infty$.
QSH and OI denotes quantum spin Hall insulator 
where spin Hall conductance is quantized and ordinary insulator respectively for $N_z=1$.
${\bar M_1}=0.016, 0.070, 0.116, 0.140, 0.216, 0.316$, and $0.416 {\rm eV}$ 
for (a), (b), (c), (d), (e), (f), and (g), respectively.
These values
correspond to those used in Figure \ref{gap_all}.
}
\label{souzu}
\end{figure}

In the following, we focus on the material parameters dependencies of $E_{\rm g}$.
For simplicity, we fix all parameters except for $\bar M_1$.
Here, we choose seven cases of $\bar M_1$ as shown in Figure \ref{souzu}.
In order to understand 
electronic properties for the corresponding seven cases we have chosen,
we show the phase diagram of the system for $N_z=1$ and $N_z=\infty$ in Figure \ref{souzu}.
In the limit for $N_z = \infty$,
the system becomes weak topological insulator
(WTI) for $\bar M_1 <  - \bar M_0/4 = 0.07 {\rm eV}$ 
while it becomes strong topological insulator (STI)
for $\bar M_1 >  - \bar M_0/4 = 0.07 {\rm eV}$.
On the other hand,
in the limit for $N_z=1$
the present system is 
QSH
for  $\bar M_1 <  - \bar M_0/2 = 0.14 {\rm eV}$,
while ordinary insulator (OI) for $\bar M_1 <  - \bar M_0/2 = 0.14 {\rm eV}$.
\begin{figure}
\centering
\includegraphics{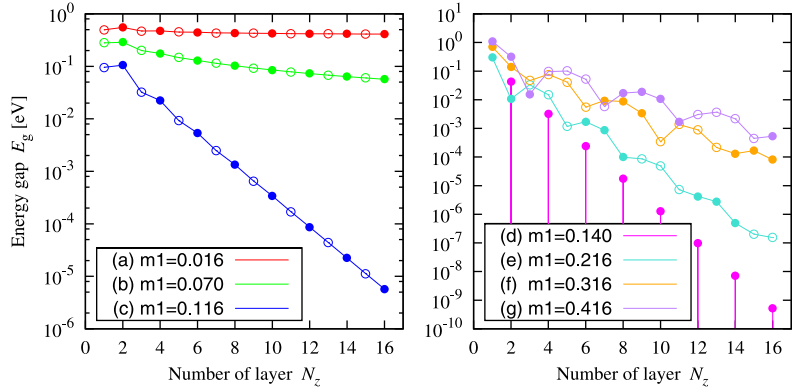}
\caption{The magnitude of energy gap $E_{\rm g}$ as a function of $N_z$ for different values of ${\bar M_1}$.
The closed (open) circle describes a two--dimensional topological invariant $\nu=0 (\nu=1)$.
At the case (d), $E_g$ vanishes for odd numbers of layers: $N_z=1,3,\cdots, 15$.
}
\label{gap_all}
\end{figure}

Figure \ref{gap_all} shows $E_{\rm g}$ for  various values of $\bar M_1$.
The curve in Figure \ref{EV} coincides with the curve (e) 
obtained for $\bar M_1 = 0.216 {\rm eV}$ in Figure \ref{gap_all},
which is the same parameter as that of Bi$_2$Se$_3$.
The curve (e)
has a three-fold periodic damped oscillation.
For $\bar M_1 = 0.016 {\rm eV}$ (case (a) in Fig. \ref{gap_all}), where the system is WTI (QSH) with $N_z=\infty$ ($N_z=1$),
$E_{\rm g}$ does not decay with the increase of $N_z$ 
since there is no gapless surface Dirac cone at $\bar{\Gamma}$-point
for $N_z=\infty$.
For $\bar M_1 = 0.070 {\rm eV}$ (case (b) in Fig. \ref{gap_all}), 
which is the transition point between WTI and STI,
where
closing of the bulk energy gap occurs at $\Gamma$--point,
$E_{\rm g}$ decreases monotonically as a function of $N_z$.
For $\bar M_1 = 0.116 {\rm eV}$ (case (c) in Fig. \ref{gap_all}),
$E_{\rm g}$ decays exponentially except for $N_z < 5$ as a function of $N_z$,
since surface Dirac cone is generated with $N_z = \infty$.
$E_{\rm g}$
has a strong oscillation for $\bar M_1 = 0.140 {\rm eV}$ (case (d) in Fig. \ref{gap_all}),
where transition between OI and QSH occurs for $N_z=1$.
$E_{\rm g}$ 
becomes exactly zero for odd numbers of $N_z$ (See Appendix).
%
%
The damped oscillation with four--fold periodicity 
appears at $\bar M_1=0.316 {\rm eV}$ (case (f) in Fig. \ref{gap_all}) and at $\bar M_1=0.416 {\rm eV}$ (case (g) in Fig. \ref{gap_all}).
When the system is QSH for $N_z=1$,
$E_{\rm g}$ decreases monotonically in the wide parameter range of $N_z$. 
On the other hand,
$E_{\rm g}$ shows a damped oscillation as a function of $N_z$
for
$\bar M_1 > 0.140 {\rm eV}$, {\it i.e.},
the system is OI 
in the thin thickness limit.
As we have seen above, 
the period of oscillation
 depends on $\bar M_1$.
It can be concluded that
the 
$N_z$ dependence of
$E_{\rm g}$ is sensitive to the material parameter $\bar M_1$.
We also show the topological invariant $\nu$ in Fig.\ref{gap_all}. The closed circle expresses non-topological phase with $\nu=0$, while the open circle expresses topological phase with $\nu=1$.
For cases with (a), (b) and (c), non-topological phase emerges for even number of layers and topological phase emerges for odd number of layers. 
Topological phase and non-topological phase appear oscillatory also for cases with (e), (f) and (g).
However, the period of oscillation becomes three or four. These results are consistent with those by Liu $et$ $al.$ \cite{liu10_oscillatory}.
Since the magnitude of energy gap becomes zero at the boundary between topological and non-topological phases, the period of oscillation of topological number $\nu$ coincides with that of the energy gap.

\section{Conclusion}

We obtain the energy spectrum of
surface states in a topological insulator based on tight--binding model.
It is clarified that
there are various types of thickness dependencies of $E_{\rm g}$.
The origin of
the dumped oscillatory behavior of $E_{\rm g}$ is partitioned into two parts.
The dumped behavior
appears when the gapless surface Dirac cone is realized in the limit of 
$N_z= \infty$.
The oscillatory behavior of $E_{\rm g}$ 
becomes prominent
when the system approaches to OI regime for $N_z=1$.

Based on these results,
we would expect various types of thin films by controlling material parameters, which could be controlled by external pressure along z-direction (c-axis) for $\rm{Bi_2Se_3}$.
Tuning of material parameters may be much more easier for optical lattices made from cold atoms \cite{Goldman_prl_neutral,Bermudez_arxiv_trap,Bermudez_arxiv_defect,Bermudez_arxiv_sweep,Goldman_prl_nonabelian,Bermudez_arxiv_honeycomb}.
If we can tune the corresponding material parameters with the case (d) in Fig.4, the strong even-odd effect is expected. In this case, various transport properties are sensitive to external fields.
%
There are several future unresolved problems.
The Anderson localizations of thin films have been recently studied \cite{hirahara10,liu11,wang11PRB} from various aspects.
It is interesting to study this problem with various types of thin films
with different electronic properties
 of topological insulator with different material parameters.

\section{Acknowledgments}
This work was supported
in part by a Grant-in-Aid for Scientific Research from MEXT
of Japan, ``Topological Quantum Phenomena" No. 22103005
and No. 22340096.

\section*{Appendix}

As shown in Fig. \ref{gap_all}, the energy gap is exactly zero for the odd number of layers, and non-zero for the even number of layers at the case (d) in Fig. \ref{souzu}. In the following, we derive this behavior of the energy gap.
The Hamiltonian for $N_z$ at $\bar \Gamma$ point reads
\begin{eqnarray}
 H_{N_z}(0,0) = F_{N_z}+G_{N_z},
\end{eqnarray}
with $4N_z\times 4N_z$ matrices $F_{N_z}$ and $G_{N_z}$ being
\begin{eqnarray}
F_{N_z}&=&\sum_{n=1}^{N_z} c^\dag_{n} {\rm diag}\left[ ( {\bar C_0}+2{\bar C_1}),\cdots,({\bar C_0}+2{\bar C_1})\right] c_{n}\\
G_{N_z}&=&\sum_{n=1}^{N_z-1} 
\left[ c^\dag_{n} H_1 c_{n+1} + \mathrm{h.c.}
\right].
\end{eqnarray}
where ${\bar M_0}+2{\bar M_1}=0$ at the case (d). The eigenvalue of $ H_{N_z}(0,0) $ equals to ${\bar C_0}+2{\bar C_1}+E^G_n$ where $E^G_n$ is the eigenvalue of $G_{N_z}$.
If matrix $G_{N_z}$ has zero eigenvalue, the energy gap closes because $G_{N_z}$ has particle hole symmetry .
We show $|G_{N_z}| \equiv \det G_{N_z}=0$  for odd number of $N_z$ at $\bar \Gamma$ point, as follows.
\begin{eqnarray}\nonumber
  |G_{N_z}|&=&\left|
 \begin{array}{cccccc}
 0 &H_1 &0& &&\\
 H_1^\dagger&0&H_1&&&\\
 0&H_1^\dagger&0&\ddots&&\\
 &&\ddots&\ddots&\ddots&\\
 &&&\ddots&0&H_1\\
 &&&&H_1^\dagger&0
 \end{array}
 \right|=\left|
 \begin{array}{cccccc}
H_1 &0&0& &&\\
0&H_1^\dagger&H_1&&&\\
 H_1^\dagger&0&0&\ddots&&\\
 &&H_1^\dagger&\ddots&\ddots&\\
 &&&\ddots&0&H_1\\
 &&&&H_1^\dagger&0
 \end{array}
 \right|\\\nonumber
 &=&\left| H_1 \right| \left|
  \begin{array}{ccccc}
H_1^\dagger&H_1&&&\\
0&0&\ddots&&\\
 &H_1^\dagger&\ddots&\ddots&\\
 &&\ddots&0&H_1\\
 &&&H_1^\dagger&0
 \end{array}
 \right|=\left| H_1\right|^2\left|
 \begin{array}{cccc}
0&H_1&&\\
H_1^\dagger&\ddots&\ddots&\\
 &\ddots&0&H_1\\
 &&H_1^\dagger&0
 \end{array}
 \right| \\
 &=&\left| H_1\right|^2 |G_{N_z-2}|,
\end{eqnarray}
and we find
 \begin{eqnarray}
\left|G_{N_z=1}\right|&=& 0,\\
\left|G_{N_z=2}\right|&=& \left|
\begin{array}{cc}
0&H_1\\
H_1^\dagger&0
 \end{array}
\right|=\left| H_1\right| ^2 \neq 0.
\end{eqnarray}
Thus the energy gap at the case (d) is exactly zero for the odd number of layers, and non-zero for the even number of layers, and oscillates strongly as a function of $N_z$.


\begin{thebibliography}{54}
\expandafter\ifx\csname natexlab\endcsname\relax\def\natexlab#1{#1}\fi
\providecommand{\bibinfo}[2]{#2}
\ifx\xfnm\relax \def\xfnm[#1]{\unskip,\space#1}\fi
\bibitem[{Hasan and Kane(2010)}]{hasan10}
\bibinfo{author}{M.~Z. Hasan}, \bibinfo{author}{C.~L. Kane},
  \bibinfo{journal}{Rev. Mod. Phys.} \bibinfo{volume}{82}
  (\bibinfo{year}{2010}) \bibinfo{pages}{3045}.
\bibitem[{{Qi} and {Zhang}(2010{\natexlab{a}})}]{qi10_QSH}
\bibinfo{author}{X.-L. {Qi}}, \bibinfo{author}{S.-C. {Zhang}},
  \bibinfo{journal}{Physics Today} \bibinfo{volume}{63}
  (\bibinfo{year}{2010}{\natexlab{a}}) \bibinfo{pages}{33}.
\bibitem[{{Qi} and {Zhang}(2010{\natexlab{b}})}]{qi10_topo}
\bibinfo{author}{X.-L. {Qi}}, \bibinfo{author}{S.-C. {Zhang}},
  \bibinfo{journal}{ArXiv:1008.2026}  (\bibinfo{year}{2010}{\natexlab{b}}).
\bibitem[{Moore(2010)}]{moore10}
\bibinfo{author}{J.~Moore}, \bibinfo{journal}{Nature} \bibinfo{volume}{464}
  (\bibinfo{year}{2010}) \bibinfo{pages}{194}.
\bibitem[{{Z. Hasan} et~al.(2011){Z. Hasan}, {Hsieh}, {Xia}, {Wray}, {Xu}, and
  {Kane}}]{hasan11}
\bibinfo{author}{M.~{Z. Hasan}}, \bibinfo{author}{D.~{Hsieh}},
  \bibinfo{author}{Y.~{Xia}}, \bibinfo{author}{L.~A. {Wray}},
  \bibinfo{author}{S.-Y. {Xu}}, \bibinfo{author}{C.~L. {Kane}},
  \bibinfo{journal}{ArXiv:1105.0396}  (\bibinfo{year}{2011}).
\bibitem[{Kane and Mele(2005{\natexlab{a}})}]{kane05QSH}
\bibinfo{author}{C.~L. Kane}, \bibinfo{author}{E.~J. Mele},
  \bibinfo{journal}{Phys. Rev. Lett.} \bibinfo{volume}{95}
  (\bibinfo{year}{2005}{\natexlab{a}}) \bibinfo{pages}{226801}.
\bibitem[{Kane and Mele(2005{\natexlab{b}})}]{kane05Z2}
\bibinfo{author}{C.~L. Kane}, \bibinfo{author}{E.~J. Mele},
  \bibinfo{journal}{Phys. Rev. Lett.} \bibinfo{volume}{95}
  (\bibinfo{year}{2005}{\natexlab{b}}) \bibinfo{pages}{146802}.
\bibitem[{Fu and Kane(2006)}]{fu06}
\bibinfo{author}{L.~Fu}, \bibinfo{author}{C.~L. Kane}, \bibinfo{journal}{Phys.
  Rev. B} \bibinfo{volume}{74} (\bibinfo{year}{2006}) \bibinfo{pages}{195312}.
\bibitem[{{Bernevig} et~al.(2006){Bernevig}, {Hughes}, and
  {Zhang}}]{bernevig06}
\bibinfo{author}{B.~A. {Bernevig}}, \bibinfo{author}{T.~L. {Hughes}},
  \bibinfo{author}{S.-C. {Zhang}}, \bibinfo{journal}{Science}
  \bibinfo{volume}{314} (\bibinfo{year}{2006}) \bibinfo{pages}{1757}.
\bibitem[{{K{\"o}nig} et~al.(2007){K{\"o}nig}, {Wiedmann}, {Br{\"u}ne}, {Roth},
  {Buhmann}, {Molenkamp}, {Qi}, and {Zhang}}]{koenig07}
\bibinfo{author}{M.~{K{\"o}nig}}, \bibinfo{author}{S.~{Wiedmann}},
  \bibinfo{author}{C.~{Br{\"u}ne}}, \bibinfo{author}{A.~{Roth}},
  \bibinfo{author}{H.~{Buhmann}}, \bibinfo{author}{L.~W. {Molenkamp}},
  \bibinfo{author}{X.-L. {Qi}}, \bibinfo{author}{S.-C. {Zhang}},
  \bibinfo{journal}{Science} \bibinfo{volume}{318} (\bibinfo{year}{2007})
  \bibinfo{pages}{766}.
\bibitem[{Roth et~al.(2009)Roth, Br{\"u}ne, Buhmann, Molenkamp, Maciejko, Qi,
  and Zhang}]{roth09}
\bibinfo{author}{A.~Roth}, \bibinfo{author}{C.~Br{\"u}ne},
  \bibinfo{author}{H.~Buhmann}, \bibinfo{author}{L.~W. Molenkamp},
  \bibinfo{author}{J.~Maciejko}, \bibinfo{author}{X.-L. Qi},
  \bibinfo{author}{S.-C. Zhang}, \bibinfo{journal}{Science}
  \bibinfo{volume}{325} (\bibinfo{year}{2009}) \bibinfo{pages}{294}.
\bibitem[{{Br{\"u}ne} et~al.(2011){Br{\"u}ne}, {Roth}, {Buhmann}, {Hankiewicz},
  {Molenkamp}, {Maciejko}, {Qi}, and {Zhang}}]{brune11}
\bibinfo{author}{C.~{Br{\"u}ne}}, \bibinfo{author}{A.~{Roth}},
  \bibinfo{author}{H.~{Buhmann}}, \bibinfo{author}{E.~M. {Hankiewicz}},
  \bibinfo{author}{L.~W. {Molenkamp}}, \bibinfo{author}{J.~{Maciejko}},
  \bibinfo{author}{X.-L. {Qi}}, \bibinfo{author}{S.-C. {Zhang}},
  \bibinfo{journal}{ArXiv:1107.0585}  (\bibinfo{year}{2011}).
\bibitem[{Fu et~al.(2007)Fu, Kane, and Mele}]{fu07_3D}
\bibinfo{author}{L.~Fu}, \bibinfo{author}{C.~L. Kane}, \bibinfo{author}{E.~J.
  Mele}, \bibinfo{journal}{Phys. Rev. Lett.} \bibinfo{volume}{98}
  (\bibinfo{year}{2007}) \bibinfo{pages}{106803}.
\bibitem[{Fu and Kane(2007)}]{fu07_inversion}
\bibinfo{author}{L.~Fu}, \bibinfo{author}{C.~L. Kane}, \bibinfo{journal}{Phys.
  Rev. B} \bibinfo{volume}{76} (\bibinfo{year}{2007}) \bibinfo{pages}{045302}.
\bibitem[{Moore and Balents(2007)}]{moore07}
\bibinfo{author}{J.~E. Moore}, \bibinfo{author}{L.~Balents},
  \bibinfo{journal}{Phys. Rev. B} \bibinfo{volume}{75} (\bibinfo{year}{2007})
  \bibinfo{pages}{121306}.
\bibitem[{Hsieh et~al.(2008)Hsieh, Qian, Wray, Xia, Hor, Cava, and
  Hasan}]{hsieh08}
\bibinfo{author}{D.~Hsieh}, \bibinfo{author}{D.~Qian},
  \bibinfo{author}{L.~Wray}, \bibinfo{author}{Y.~Xia}, \bibinfo{author}{Y.~S.
  Hor}, \bibinfo{author}{R.~J. Cava}, \bibinfo{author}{M.~Z. Hasan},
  \bibinfo{journal}{Nature} \bibinfo{volume}{452} (\bibinfo{year}{2008})
  \bibinfo{pages}{970--974}.
\bibitem[{Zhang et~al.(2009)Zhang, Liu, Qi, Dai, Fang, and Zhang}]{zhang09}
\bibinfo{author}{H.~Zhang}, \bibinfo{author}{C.-X. Liu}, \bibinfo{author}{X.-L.
  Qi}, \bibinfo{author}{X.~Dai}, \bibinfo{author}{Z.~Fang},
  \bibinfo{author}{S.-C. Zhang}, \bibinfo{journal}{Nature Phys.}
  \bibinfo{volume}{5} (\bibinfo{year}{2009}) \bibinfo{pages}{438--442}.
\bibitem[{Liu et~al.(2010)Liu, Qi, Zhang, Dai, Fang, and Zhang}]{liu10_model}
\bibinfo{author}{C.-X. Liu}, \bibinfo{author}{X.-L. Qi},
  \bibinfo{author}{H.~Zhang}, \bibinfo{author}{X.~Dai},
  \bibinfo{author}{Z.~Fang}, \bibinfo{author}{S.-C. Zhang},
  \bibinfo{journal}{Phys. Rev. B} \bibinfo{volume}{82} (\bibinfo{year}{2010})
  \bibinfo{pages}{045122}.
\bibitem[{Xia et~al.(2009)Xia, Qian, Hsieh, Wray, Pal, Lin, Bansil, Grauer,
  Hor, Cava, and Hasan}]{xia09}
\bibinfo{author}{Y.~Xia}, \bibinfo{author}{D.~Qian},
  \bibinfo{author}{D.~Hsieh}, \bibinfo{author}{L.~Wray},
  \bibinfo{author}{A.~Pal}, \bibinfo{author}{H.~Lin},
  \bibinfo{author}{A.~Bansil}, \bibinfo{author}{D.~Grauer},
  \bibinfo{author}{Y.~S. Hor}, \bibinfo{author}{R.~J. Cava},
  \bibinfo{author}{M.~Z. Hasan}, \bibinfo{journal}{Nature Phys.}
  \bibinfo{volume}{5} (\bibinfo{year}{2009}) \bibinfo{pages}{398--402}.
\bibitem[{Chen et~al.(2009)Chen, Analytis, Chu, Liu, Mo, Qi, Zhang, Lu, Dai,
  Fang, Zhang, Fisher, Hussain, and Shen}]{chen09}
\bibinfo{author}{Y.~L. Chen}, \bibinfo{author}{J.~G. Analytis},
  \bibinfo{author}{J.-H. Chu}, \bibinfo{author}{Z.~K. Liu},
  \bibinfo{author}{S.-K. Mo}, \bibinfo{author}{X.-L. Qi},
  \bibinfo{author}{H.~J. Zhang}, \bibinfo{author}{D.~H. Lu},
  \bibinfo{author}{X.~Dai}, \bibinfo{author}{Z.~Fang}, \bibinfo{author}{S.-C.
  Zhang}, \bibinfo{author}{I.~R. Fisher}, \bibinfo{author}{Z.~Hussain},
  \bibinfo{author}{Z.-X. Shen}, \bibinfo{journal}{Science}
  \bibinfo{volume}{325} (\bibinfo{year}{2009}) \bibinfo{pages}{178}.
\bibitem[{Yan et~al.(2010)Yan, Liu, Zhang, Yam, Qi, Frauenheim, and
  Zhang}]{yan10}
\bibinfo{author}{B.~Yan}, \bibinfo{author}{C.-X. Liu}, \bibinfo{author}{H.-J.
  Zhang}, \bibinfo{author}{C.-Y. Yam}, \bibinfo{author}{X.-L. Qi},
  \bibinfo{author}{T.~Frauenheim}, \bibinfo{author}{S.-C. Zhang},
  \bibinfo{journal}{Europhys. Lett.} \bibinfo{volume}{90}
  (\bibinfo{year}{2010}) \bibinfo{pages}{37002}.
\bibitem[{Lin et~al.(2010)Lin, Markiewicz, Wray, Fu, Hasan, and Bansil}]{lin10}
\bibinfo{author}{H.~Lin}, \bibinfo{author}{R.~S. Markiewicz},
  \bibinfo{author}{L.~A. Wray}, \bibinfo{author}{L.~Fu}, \bibinfo{author}{M.~Z.
  Hasan}, \bibinfo{author}{A.~Bansil}, \bibinfo{journal}{Phys. Rev. Lett.}
  \bibinfo{volume}{105} (\bibinfo{year}{2010}) \bibinfo{pages}{036404}.
\bibitem[{Sato et~al.(2010)Sato, Segawa, Guo, Sugawara, Souma, Takahashi, and
  Ando}]{sato10}
\bibinfo{author}{T.~Sato}, \bibinfo{author}{K.~Segawa},
  \bibinfo{author}{H.~Guo}, \bibinfo{author}{K.~Sugawara},
  \bibinfo{author}{S.~Souma}, \bibinfo{author}{T.~Takahashi},
  \bibinfo{author}{Y.~Ando}, \bibinfo{journal}{Phys. Rev. Lett.}
  \bibinfo{volume}{105} (\bibinfo{year}{2010}) \bibinfo{pages}{136802}.
\bibitem[{Chen et~al.(2011)Chen, Gong, Duan, Zhu, Chu, Walsh, Yao, Ma, and
  Wei}]{chen11}
\bibinfo{author}{S.~Chen}, \bibinfo{author}{X.~G. Gong}, \bibinfo{author}{C.-G.
  Duan}, \bibinfo{author}{Z.-Q. Zhu}, \bibinfo{author}{J.-H. Chu},
  \bibinfo{author}{A.~Walsh}, \bibinfo{author}{Y.-G. Yao},
  \bibinfo{author}{J.~Ma}, \bibinfo{author}{S.-H. Wei}, \bibinfo{journal}{Phys.
  Rev. B} \bibinfo{volume}{83} (\bibinfo{year}{2011}) \bibinfo{pages}{245202}.
\bibitem[{{Wang} et~al.(2011){Wang}, {Lin}, {Das}, {Hasan}, and
  {Bansil}}]{wang11arXiv}
\bibinfo{author}{Y.~J. {Wang}}, \bibinfo{author}{H.~{Lin}},
  \bibinfo{author}{T.~{Das}}, \bibinfo{author}{M.~Z. {Hasan}},
  \bibinfo{author}{A.~{Bansil}}, \bibinfo{journal}{ArXiv:1106.3316}
  (\bibinfo{year}{2011}).
\bibitem[{Tanaka et~al.(2009)Tanaka, Yokoyama, and Nagaosa}]{TYN09}
\bibinfo{author}{Y.~Tanaka}, \bibinfo{author}{T.~Yokoyama},
  \bibinfo{author}{N.~Nagaosa}, \bibinfo{journal}{Phys. Rev. Lett.}
  \bibinfo{volume}{103} (\bibinfo{year}{2009}) \bibinfo{pages}{107002}.
\bibitem[{Linder et~al.(2010{\natexlab{a}})Linder, Tanaka, Yokoyama, Sudbo, and
  Nagaosa}]{Linder10a}
\bibinfo{author}{J.~Linder}, \bibinfo{author}{Y.~Tanaka},
  \bibinfo{author}{T.~Yokoyama}, \bibinfo{author}{A.~Sudbo},
  \bibinfo{author}{N.~Nagaosa}, \bibinfo{journal}{Phys. Rev. Lett.}
  \bibinfo{volume}{104} (\bibinfo{year}{2010}{\natexlab{a}})
  \bibinfo{pages}{067001}.
\bibitem[{Linder et~al.(2010{\natexlab{b}})Linder, Tanaka, Yokoyama, Sudbo, and
  Nagaosa}]{Linder10b2}
\bibinfo{author}{J.~Linder}, \bibinfo{author}{Y.~Tanaka},
  \bibinfo{author}{T.~Yokoyama}, \bibinfo{author}{A.~Sudbo},
  \bibinfo{author}{N.~Nagaosa}, \bibinfo{journal}{Phys. Rev. B}
  \bibinfo{volume}{81} (\bibinfo{year}{2010}{\natexlab{b}})
  \bibinfo{pages}{184525}.
\bibitem[{Yokoyama et~al.(2010)Yokoyama, Tanaka, and Nagaosa}]{Yokoyama}
\bibinfo{author}{T.~Yokoyama}, \bibinfo{author}{Y.~Tanaka},
  \bibinfo{author}{N.~Nagaosa}, \bibinfo{journal}{Phys. Rev. B}
  \bibinfo{volume}{81} (\bibinfo{year}{2010}) \bibinfo{pages}{121401}.
\bibitem[{Hyde et~al.(1974)Hyde, Beale, Spain, and Woollam}]{hyde74}
\bibinfo{author}{G.~R. Hyde}, \bibinfo{author}{H.~A. Beale},
  \bibinfo{author}{I.~L. Spain}, \bibinfo{author}{J.~A. Woollam},
  \bibinfo{journal}{J. Phys. Chem. Solids} \bibinfo{volume}{35}
  (\bibinfo{year}{1974}) \bibinfo{pages}{1719--1728}.
\bibitem[{Taskin and Ando(2009)}]{taskin09}
\bibinfo{author}{A.~A. Taskin}, \bibinfo{author}{Y.~Ando},
  \bibinfo{journal}{Phys. Rev. B} \bibinfo{volume}{80} (\bibinfo{year}{2009})
  \bibinfo{pages}{085303}.
\bibitem[{Checkelsky et~al.(2009)Checkelsky, Hor, Liu, Qu, Cava, and
  Ong}]{checkelsky09}
\bibinfo{author}{J.~G. Checkelsky}, \bibinfo{author}{Y.~S. Hor},
  \bibinfo{author}{M.-H. Liu}, \bibinfo{author}{D.-X. Qu},
  \bibinfo{author}{R.~J. Cava}, \bibinfo{author}{N.~P. Ong},
  \bibinfo{journal}{Phys. Rev. Lett.} \bibinfo{volume}{103}
  (\bibinfo{year}{2009}) \bibinfo{pages}{246601}.
\bibitem[{Hsieh et~al.(2009)Hsieh, Xia, Qian, Wray, Dil, Meier, Osterwalder,
  Patthey, Checkelsky, Ong, Fedorov, Lin, Bansil, Grauer, Hor, Cava, and
  Hasan}]{hsieh09}
\bibinfo{author}{D.~Hsieh}, \bibinfo{author}{Y.~Xia},
  \bibinfo{author}{D.~Qian}, \bibinfo{author}{L.~Wray}, \bibinfo{author}{J.~H.
  Dil}, \bibinfo{author}{F.~Meier}, \bibinfo{author}{J.~Osterwalder},
  \bibinfo{author}{L.~Patthey}, \bibinfo{author}{J.~G. Checkelsky},
  \bibinfo{author}{N.~P. Ong}, \bibinfo{author}{A.~V. Fedorov},
  \bibinfo{author}{H.~Lin}, \bibinfo{author}{A.~Bansil},
  \bibinfo{author}{D.~Grauer}, \bibinfo{author}{Y.~S. Hor},
  \bibinfo{author}{R.~J. Cava}, \bibinfo{author}{M.~Z. Hasan},
  \bibinfo{journal}{Nature} \bibinfo{volume}{460} (\bibinfo{year}{2009})
  \bibinfo{pages}{1101--1105}.
\bibitem[{Chen et~al.(2010)Chen, Qin, Yang, Liu, Guan, Qu, Zhang, Shi, Xie,
  Yang, Wu, Li, and Lu}]{chen10PRL}
\bibinfo{author}{J.~Chen}, \bibinfo{author}{H.~J. Qin},
  \bibinfo{author}{F.~Yang}, \bibinfo{author}{J.~Liu},
  \bibinfo{author}{T.~Guan}, \bibinfo{author}{F.~M. Qu}, \bibinfo{author}{G.~H.
  Zhang}, \bibinfo{author}{J.~R. Shi}, \bibinfo{author}{X.~C. Xie},
  \bibinfo{author}{C.~L. Yang}, \bibinfo{author}{K.~H. Wu},
  \bibinfo{author}{Y.~Q. Li}, \bibinfo{author}{L.~Lu}, \bibinfo{journal}{Phys.
  Rev. Lett.} \bibinfo{volume}{105} (\bibinfo{year}{2010})
  \bibinfo{pages}{176602}.
\bibitem[{Chen et~al.(2011)Chen, He, Wu, Ji, Lu, Shi, Smet, and Li}]{chen10PRB}
\bibinfo{author}{J.~Chen}, \bibinfo{author}{X.~Y. He}, \bibinfo{author}{K.~H.
  Wu}, \bibinfo{author}{Z.~Q. Ji}, \bibinfo{author}{L.~Lu},
  \bibinfo{author}{J.~R. Shi}, \bibinfo{author}{J.~H. Smet},
  \bibinfo{author}{Y.~Q. Li}, \bibinfo{journal}{Phys. Rev. B}
  \bibinfo{volume}{83} (\bibinfo{year}{2011}) \bibinfo{pages}{241304}.
\bibitem[{{Zhang} et~al.(2010){Zhang}, {He}, {Chang}, {Song}, {Wang}, {Chen},
  {Jia}, {Fang}, {Dai}, {Shan}, {Shen}, {Niu}, {Qi}, {Zhang}, {Ma}, and
  {Xue}}]{zhang10}
\bibinfo{author}{Y.~{Zhang}}, \bibinfo{author}{K.~{He}}, \bibinfo{author}{C.-Z.
  {Chang}}, \bibinfo{author}{C.-L. {Song}}, \bibinfo{author}{L.-L. {Wang}},
  \bibinfo{author}{X.~{Chen}}, \bibinfo{author}{J.-F. {Jia}},
  \bibinfo{author}{Z.~{Fang}}, \bibinfo{author}{X.~{Dai}},
  \bibinfo{author}{W.-Y. {Shan}}, \bibinfo{author}{S.-Q. {Shen}},
  \bibinfo{author}{Q.~{Niu}}, \bibinfo{author}{X.-L. {Qi}},
  \bibinfo{author}{S.-C. {Zhang}}, \bibinfo{author}{X.-C. {Ma}},
  \bibinfo{author}{Q.-K. {Xue}}, \bibinfo{journal}{Nature Phys.}
  \bibinfo{volume}{6} (\bibinfo{year}{2010}) \bibinfo{pages}{712}.
\bibitem[{Sakamoto et~al.(2010)Sakamoto, Hirahara, Miyazaki, Kimura, and
  Hasegawa}]{sakamoto10}
\bibinfo{author}{Y.~Sakamoto}, \bibinfo{author}{T.~Hirahara},
  \bibinfo{author}{H.~Miyazaki}, \bibinfo{author}{S.-i. Kimura},
  \bibinfo{author}{S.~Hasegawa}, \bibinfo{journal}{Phys. Rev. B}
  \bibinfo{volume}{81} (\bibinfo{year}{2010}) \bibinfo{pages}{165432}.
\bibitem[{Hirahara et~al.(2010)Hirahara, Sakamoto, Takeichi, Miyazaki, Kimura,
  Matsuda, Kakizaki, and Hasegawa}]{hirahara10}
\bibinfo{author}{T.~Hirahara}, \bibinfo{author}{Y.~Sakamoto},
  \bibinfo{author}{Y.~Takeichi}, \bibinfo{author}{H.~Miyazaki},
  \bibinfo{author}{S.-i. Kimura}, \bibinfo{author}{I.~Matsuda},
  \bibinfo{author}{A.~Kakizaki}, \bibinfo{author}{S.~Hasegawa},
  \bibinfo{journal}{Phys. Rev. B} \bibinfo{volume}{82} (\bibinfo{year}{2010})
  \bibinfo{pages}{155309}.
\bibitem[{Linder et~al.(2009)Linder, Yokoyama, and Sudb\o{}}]{linder09}
\bibinfo{author}{J.~Linder}, \bibinfo{author}{T.~Yokoyama},
  \bibinfo{author}{A.~Sudb\o{}}, \bibinfo{journal}{Phys. Rev. B}
  \bibinfo{volume}{80} (\bibinfo{year}{2009}) \bibinfo{pages}{205401}.
\bibitem[{Lu et~al.(2010)Lu, Shan, Yao, Niu, and Shen}]{lu10}
\bibinfo{author}{H.-Z. Lu}, \bibinfo{author}{W.-Y. Shan},
  \bibinfo{author}{W.~Yao}, \bibinfo{author}{Q.~Niu}, \bibinfo{author}{S.-Q.
  Shen}, \bibinfo{journal}{Phys. Rev. B} \bibinfo{volume}{81}
  (\bibinfo{year}{2010}) \bibinfo{pages}{115407}.
\bibitem[{{Shan} et~al.(2010){Shan}, {Lu}, and {Shen}}]{shan10}
\bibinfo{author}{W.-Y. {Shan}}, \bibinfo{author}{H.-Z. {Lu}},
  \bibinfo{author}{S.-Q. {Shen}}, \bibinfo{journal}{New J. Phys.}
  \bibinfo{volume}{12} (\bibinfo{year}{2010}) \bibinfo{pages}{043048}.
\bibitem[{Liu et~al.(2010)Liu, Zhang, Yan, Qi, Frauenheim, Dai, Fang, and
  Zhang}]{liu10_oscillatory}
\bibinfo{author}{C.-X. Liu}, \bibinfo{author}{H.~Zhang},
  \bibinfo{author}{B.~Yan}, \bibinfo{author}{X.-L. Qi},
  \bibinfo{author}{T.~Frauenheim}, \bibinfo{author}{X.~Dai},
  \bibinfo{author}{Z.~Fang}, \bibinfo{author}{S.-C. Zhang},
  \bibinfo{journal}{Phys. Rev. B} \bibinfo{volume}{81} (\bibinfo{year}{2010})
  \bibinfo{pages}{041307}.
\bibitem[{Park et~al.(2010)Park, Heremans, Scarola, and Minic}]{park10}
\bibinfo{author}{K.~Park}, \bibinfo{author}{J.~J. Heremans},
  \bibinfo{author}{V.~W. Scarola}, \bibinfo{author}{D.~Minic},
  \bibinfo{journal}{Phys. Rev. Lett.} \bibinfo{volume}{105}
  (\bibinfo{year}{2010}) \bibinfo{pages}{186801}.
\bibitem[{Yazyev et~al.(2010)Yazyev, Moore, and Louie}]{yazyev10}
\bibinfo{author}{O.~V. Yazyev}, \bibinfo{author}{J.~E. Moore},
  \bibinfo{author}{S.~G. Louie}, \bibinfo{journal}{Phys. Rev. Lett.}
  \bibinfo{volume}{105} (\bibinfo{year}{2010}) \bibinfo{pages}{266806}.
\bibitem[{Jin et~al.(2011)Jin, Song, and Freeman}]{jin11}
\bibinfo{author}{H.~Jin}, \bibinfo{author}{J.-H. Song}, \bibinfo{author}{A.~J.
  Freeman}, \bibinfo{journal}{Phys. Rev. B} \bibinfo{volume}{83}
  (\bibinfo{year}{2011}) \bibinfo{pages}{125319}.
\bibitem[{Chang et~al.(2011)Chang, Register, Banerjee, and Sahu}]{chang11}
\bibinfo{author}{J.~Chang}, \bibinfo{author}{L.~F. Register},
  \bibinfo{author}{S.~K. Banerjee}, \bibinfo{author}{B.~Sahu},
  \bibinfo{journal}{Phys. Rev. B} \bibinfo{volume}{83} (\bibinfo{year}{2011})
  \bibinfo{pages}{235108}.
\bibitem[{Goldman et~al.(2010)Goldman, Satija, Nikolic, Bermudez,
  Martin-Delgado, Lewenstein, and Spielman}]{Goldman_prl_neutral}
\bibinfo{author}{N.~Goldman}, \bibinfo{author}{I.~Satija},
  \bibinfo{author}{P.~Nikolic}, \bibinfo{author}{A.~Bermudez},
  \bibinfo{author}{M.~A. Martin-Delgado}, \bibinfo{author}{M.~Lewenstein},
  \bibinfo{author}{I.~B. Spielman}, \bibinfo{journal}{Phys. Rev. Lett.}
  \bibinfo{volume}{105} (\bibinfo{year}{2010}) \bibinfo{pages}{255302}.
\bibitem[{{Bermudez} et~al.(2010){Bermudez}, {Martin-Delgado}, and
  {Porras}}]{Bermudez_arxiv_trap}
\bibinfo{author}{A.~{Bermudez}}, \bibinfo{author}{M.~A. {Martin-Delgado}},
  \bibinfo{author}{D.~{Porras}}, \bibinfo{journal}{New Journal of Physics}
  \bibinfo{volume}{12} (\bibinfo{year}{2010}) \bibinfo{pages}{123016}.
\bibitem[{{Bermudez} et~al.(2009){Bermudez}, {Patan{\`e}}, {Amico}, and
  {Martin-Delgado}}]{Bermudez_arxiv_defect}
\bibinfo{author}{A.~{Bermudez}}, \bibinfo{author}{D.~{Patan{\`e}}},
  \bibinfo{author}{L.~{Amico}}, \bibinfo{author}{M.~A. {Martin-Delgado}},
  \bibinfo{journal}{Physical Review Letters} \bibinfo{volume}{102}
  (\bibinfo{year}{2009}) \bibinfo{pages}{135702}.
\bibitem[{{Bermudez} et~al.(2010){Bermudez}, {Amico}, and
  {Martin-Delgado}}]{Bermudez_arxiv_sweep}
\bibinfo{author}{A.~{Bermudez}}, \bibinfo{author}{L.~{Amico}},
  \bibinfo{author}{M.~A. {Martin-Delgado}}, \bibinfo{journal}{New Journal of
  Physics} \bibinfo{volume}{12} (\bibinfo{year}{2010}) \bibinfo{pages}{055014}.
\bibitem[{Goldman et~al.(2009)Goldman, Kubasiak, Bermudez, Gaspard, Lewenstein,
  and Martin-Delgado}]{Goldman_prl_nonabelian}
\bibinfo{author}{N.~Goldman}, \bibinfo{author}{A.~Kubasiak},
  \bibinfo{author}{A.~Bermudez}, \bibinfo{author}{P.~Gaspard},
  \bibinfo{author}{M.~Lewenstein}, \bibinfo{author}{M.~A. Martin-Delgado},
  \bibinfo{journal}{Phys. Rev. Lett.} \bibinfo{volume}{103}
  (\bibinfo{year}{2009}) \bibinfo{pages}{035301}.
\bibitem[{{Bermudez} et~al.(2010){Bermudez}, {Goldman}, {Kubasiak},
  {Lewenstein}, and {Martin-Delgado}}]{Bermudez_arxiv_honeycomb}
\bibinfo{author}{A.~{Bermudez}}, \bibinfo{author}{N.~{Goldman}},
  \bibinfo{author}{A.~{Kubasiak}}, \bibinfo{author}{M.~{Lewenstein}},
  \bibinfo{author}{M.~A. {Martin-Delgado}}, \bibinfo{journal}{New Journal of
  Physics} \bibinfo{volume}{12} (\bibinfo{year}{2010}) \bibinfo{pages}{033041}.
\bibitem[{Liu et~al.(2011)Liu, Chang, Zhang, Zhang, Ruan, He, Wang, Chen, Jia,
  Zhang, Xue, Ma, and Wang}]{liu11}
\bibinfo{author}{M.~Liu}, \bibinfo{author}{C.-Z. Chang},
  \bibinfo{author}{Z.~Zhang}, \bibinfo{author}{Y.~Zhang},
  \bibinfo{author}{W.~Ruan}, \bibinfo{author}{K.~He}, \bibinfo{author}{L.-l.
  Wang}, \bibinfo{author}{X.~Chen}, \bibinfo{author}{J.-F. Jia},
  \bibinfo{author}{S.-C. Zhang}, \bibinfo{author}{Q.-K. Xue},
  \bibinfo{author}{X.~Ma}, \bibinfo{author}{Y.~Wang}, \bibinfo{journal}{Phys.
  Rev. B} \bibinfo{volume}{83} (\bibinfo{year}{2011}) \bibinfo{pages}{165440}.
\bibitem[{Wang et~al.(2011)Wang, DaSilva, Chang, He, Jain, Samarth, Ma, Xue,
  and Chan}]{wang11PRB}
\bibinfo{author}{J.~Wang}, \bibinfo{author}{A.~M. DaSilva},
  \bibinfo{author}{C.-Z. Chang}, \bibinfo{author}{K.~He},
  \bibinfo{author}{J.~K. Jain}, \bibinfo{author}{N.~Samarth},
  \bibinfo{author}{X.-C. Ma}, \bibinfo{author}{Q.-K. Xue},
  \bibinfo{author}{M.~H.~W. Chan}, \bibinfo{journal}{Phys. Rev. B}
  \bibinfo{volume}{83} (\bibinfo{year}{2011}) \bibinfo{pages}{245438}.

\end{thebibliography}

\end{document}